\documentclass[a4paper]{article}

\usepackage{INTERSPEECH2019}

\title{Exploiting semi-supervised training through a dropout regularization in end-to-end speech recognition}
\name{Subhadeep Dey$^1$, Petr Motlicek$^1$, Trung Bui$^2$ and Franck Dernoncourt$^2$}
\address{
  $^1$Idiap Research Institute \\
  $^2$Adobe Research}
\email{sdey@idiap.ch, petr.motlicek@idiap.ch, bui@adobe.com, dernonco@adobe.com}

\begin{document}

\maketitle
\begin{abstract}

In this paper, we explore various approaches for semi-supervised learning in an end-to-end automatic speech recognition (ASR) framework. The first step in our approach involves training a seed model on the limited amount of labelled data. Additional unlabelled speech data is employed through a data-selection mechanism to obtain the best hypothesized output, further used to retrain the seed model. However, uncertainties of the model may not be well captured with a single hypothesis. As opposed to this technique, we apply a dropout mechanism to capture the uncertainty by obtaining multiple hypothesized text transcripts of an speech recording. We assume that the diversity of automatically generated transcripts for an utterance will implicitly increase the reliability of the model. Finally, the data-selection process is also applied on these hypothesized transcripts to reduce the uncertainty. Experiments on freely-available TEDLIUM corpus and proprietary Adobe's internal dataset show that the proposed approach significantly reduces ASR errors, compared to the baseline model.

\end{abstract}
\section{Introduction}
\label{sec:intro}


State-of-the-art approaches in automatic speech recognition (ASR) exploit the powerful discriminative capability of deep neural networks (DNN) for acoustic modelling~\cite{hinton2012deep, deng2013new, povey2016purely}. 
The current ASR advancements offer 
low error-rates, making the systems applicable for commercialization. 
In the past few years, sequence level optimization algorithms, such as lattice free maximum mutual information (LF-MMI) and end-to-end 
frameworks, have been adopted over the frame level discrimination approaches (like hybrid-DNN~\cite{motlicek2015exploiting})~\cite{kim2017joint, povey2016purely, watanabe2018espnet}. As opposed to LF-MMI, the end-to-end approaches do not require 
the creation of lexicon or decision trees for training. End-to-end sequence classification approaches such as connectionist temporal classification (CTC) and encoder-decoder frameworks 
have been successfully applied in ASR~\cite{watanabe2018espnet, miao2015eesen}. However, the end-to-end ASR requires large amount of training data to optimize the network, 
as the model needs to automatically learn the mapping from the acoustic features to text-transcripts.  



Another interesting concept is a semi-supervised learning. 
Our objective is to exploit (relatively) large amount of unlabelled data when building an end-to-end ASR system~\cite{manohar2018semi, karita2018semi}. 
This 
scenario is attractive to wide range of applications, such as low-resource speech recognition and computer vision, where unsupervised data is abundant but obtaining labels is costly~\cite{lee2013pseudo}. 
%
%
Various approaches to semi-supervised learning have been proposed in the literature~\cite{walker2017semi, karita2018semi, manohar2018semi}. A typical approach involves first training an initial seed-model on the limited amount of supervised data. The seed-model is applied on the unsupervised data to automatically generate the transcripts~\cite{lee2013pseudo,walker2017semi, bachman2014learning}. 
As 
automatically generated transcripts may be 
erroneous, 
a data-selection mechanism is applied to filter-out  the 
low confident speech utterances. 
In~\cite{walker2017semi}, the utterance-level confidences are obtained to post-process the one best hypotheses. As opposed to using only 1-best hypothesized transcripts, the whole decoding-lattice is used in~\cite{manohar2018semi} as a supervision output.  For end-to-end ASR,~\cite{karita2018semi} has recently explored an approach exploiting unpaired text and audio data. This technique proposes to extract intermediate hidden representation of speech and text data with a shared encoder network. However, this approach requires text data from the target domain 
which may not be practically available during the training stage. 

In this paper, we explore 
a data-selection mechanism for semi-supervised learning of end-to-end ASR, as it has shown to be a promising approach in various applications, such as image, text, and speech, and it has not been well explored for the end-to-end ASR. 
For data selection, we explore 
two confidence based measures, namely, (i) utterance-based decoding confidence, and (ii) entropy-based confidence. We hypothesize that these 
measures indicate the reliability of automatically generated transcripts using end-to-end ASR, given the speech recording. 
In the proposed approach, the $N$-best hypothesized 
transcripts (filtered using the confidence measures) are used to further retrain the seed-model.

Further, this paper also explores the application of dropout mechanism for augmenting the $N$-best hypothesized text. Dropout is usually employed in the conventional ASR during training as a regularizer~\cite{srivastava14a}, while 
during inference, 
the dropout is not applied. In~\cite{Vyas_ICASSP2019_2019, gal2016dropout}, dropout was applied for semi-supervised learning for characterizing the uncertainties of the DNN. Motivated by these evidences, we propose to exploit the dropout mechanism to augment the $N$-best list as follows: During the decoding of an utterance, the dropout mechanism is employed to output 1-best transcripts. The dropout is applied on the same utterance multiple times to obtain different versions of transcripts. 
We hypothesize that the diversity of decoded outputs for any utterance can localize the uncertainties of the model. Experiments are performed on the publicly-available TEDLIUM corpus and proprietary Adobe’s internal dataset. The results indicate that the proposed approach allows to efficiently 
exploit unlabelled data, leading to significant increase in ASR performance. 

This paper is organized as follows. The baseline end-to-end ASR approach is described in Section~\ref{sec:end_end}. The semi-supervised training 
is described in Section~\ref{sec:semi_supervised}. The experimental setup and results are described in Sections~\ref{sec:experimental_setup} and~\ref{sec:results} respectively. Finally, the paper is concluded in Section~\ref{sec:conclusions}.





\section{End-to-end ASR}
\label{sec:end_end}

\begin{figure}[!t]
    \centering
    \includegraphics[scale=.4]{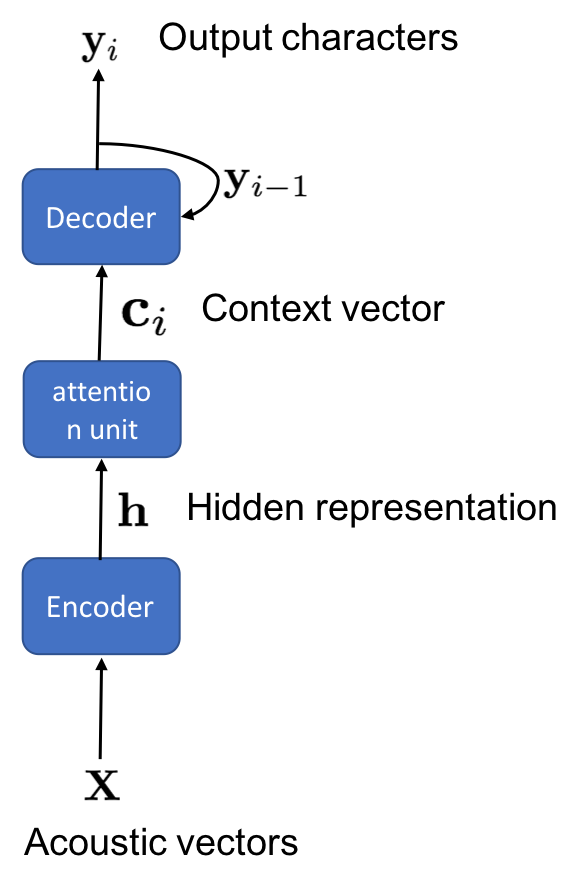}
    \caption{Architecture for encoder decoder network for end-to-end ASR.}
    \label{fig:enc-dec-baseline}
  \vspace{-4mm} 
\end{figure}

The state-of-the-art based ASR (chain model) requires  a 
lexicon and 
alignments, usually generated with respect to 
context-dependent tri-phonetic states~\cite{povey2016purely}. An alternative technique, referred to as end-to-end, aims to learn the mapping from acoustic features to text directly without the need of intermediate steps. Recently, various end-to-end sequence-to-sequence approaches, such as encoder-decoder model, have been successfully applied to ASR~\cite{watanabe2018espnet, miao2015eesen}. The basic components of the encoder-decoder model are illustrated in Figure~\ref{fig:enc-dec-baseline} and are described below: 

\begin{itemize}
    \item $\textbf{Encoder}$: The purpose of the encoder is to produce hidden representation of the utterance $\mathbf{X}$ = \{$\mathbf{x}_1$, $\mathbf{x}_2$, $\cdots$, $\mathbf{x}_T$\}, as represented by \{$\mathbf{h}_1$, $\mathbf{h}_2$, $\cdots$, $\mathbf{h}_L$\}, where L $\leq$ T. Typically, the encoder consists of a few convolutional neural network (CNN) layers and a few layers of bidirectional long short-term memory (BLSTM).   
    
    \item $\textbf{Attention unit}$: The attention unit takes as input a sequence of features and estimates the relative importance of each feature vector. The attention unit computes a context vector ($\mathbf{c}_i$) for $i^{th}$ output unit.
    
    \item $\textbf{Decoder}$: The context vector is used by the decoder unit to predict a character unit. The decoder also uses the previous decoded output character to infer current character. Training such an end-to-end ASR is performed by optimizing 
        the following loss function:
    \begin{equation}
     \label{eqn:loss_enc_dec}
        L_{loss} = -\log p(C|\mathbf{X}),
    \end{equation}
\noindent    where $C$ is the character sequence corresponding to the utterance $\mathbf{X}$. 
\end{itemize}

A connectionalist temporal classification (CTC) based loss is also combined with the objective function of Equation~\ref{eqn:loss_enc_dec} for training the network. During decoding, scores from the CTC and encoder-decoder as the acoustic models are combined using a beam-search algorithm. Furthermore, a shallow fusion of language model (LM) with the acoustic-model scores are applied to obtain text-transcripts. Further details of the end-to-end ASR  can be found in~\cite{kim2017joint, watanabe2018espnet, hori2018end}. 

\section{Semi supervised learning}
\label{sec:semi_supervised}

The end-to-end ASR is typically trained with a large amount (at least$\sim$100 hours) of labelled data~\cite{hori2018end}. However in a semi-supervised setting, it is assumed that only a small amount 
of supervised data ($\sim$10 to 15 hours) is available for training in addition to 
a large amount of untranscribed audio 
for the target domain. Estimating parameters of the end-to-end model on the limited data may not lead to a reliable solution. In this paper, we exploit publicly available data (source domain) with relatively large 
amount of speech recordings for estimating the parameters of the model. This ASR is referred to as source domain model. The parameters of the model are 
then adapted 
using the limited amount of transcribed data from the target domain. The adapted end-to-end ASR is finally used as the seed-model for exploiting the unsupervised data for further 
retraining. 

In the past, various approaches have been explored for unsupervised model adaptation~\cite{walker2017semi, karita2018semi, manohar2018semi, lee2013pseudo, bachman2014learning}. Most of the approaches rely on data-selection process for bootstrapping the model with additional labelled data selected based 
on high confidence predictions. Process of data-selection  has not been well explored for end-to-end ASR. In this paper, we explore data-selection approach using, (i) utterance-level decoding-scores, and (ii) entropy based confidence measures. These 
confidence-measures are then applied  for selecting highly reliable utterances.


The decoding-scores are obtained 
using the posterior probabilities of an utterance given the acoustic features, as a result of 
the beam-search process. Decoding-score for each $N$-best text-transcript can be generated by the ASR. Utterance-based decoding scores can be finally compared to a predefined threshold to perform a data selection. 


Furthermore, we apply entropy as a criteria to filter out utterances from the hypothesized ASR outputs. The entropy of an utterance measures the amount of uncertainty of the model. We hypothesize that entropy of the utterance is 
well  correlated  with the performance of the ASR system. The entropy of an utterance is computed as follows: The posterior probabilities of the character units are obtained by forward-pass of the model. The entropy ($\textbf{H}_u$) is then:  


\begin{equation}
    \mathbf{H}_u = - \frac{1}{T}  \sum_c^{C} p(c|\mathbf{X}) \log(p(c|\mathbf{X})),
\end{equation}

\noindent where $C$ is the number of character outputs, $p$($c$$|$$\mathbf{X}$) is the posterior probability of the $c^{th}$ character unit given the acoustic features ($\mathbf{X}$). 

We also propose to localize the uncertainty of the end-to-end ASR by applying dropout mechanism. This method is motivated by the recent advances of DNN for measuring the reliability of the model~\cite{Vyas_ICASSP2019_2019, gal2016dropout}. The conventional methods do not use dropouts during the decoding time. 
As opposed to this approach, sampling from the DNN weight distribution is done by applying the dropout. The proposed approach is as follows:

\begin{enumerate}
\item Dropout: apply dropout during the inference to obtain $1$-best hypothesized transcript 
\item Data-selection: augment the adaptation data with this utterance if the entropy or the decoding-score is above a threshold
\item Repeat steps 1 and 2 for $N$ times 
\end{enumerate} 

The above steps are applied to all the utterances of the unsupervised dataset. 



\section{Experimental Setup}
\label{sec:experimental_setup}

In this section, the experimental setup for the semi-supervised ASR is detailed. Experiments are performed on TEDLIUM and Adobe (internal) datasets as the target domain data. 

\begin {table}[t!]
\caption{Training, adaptation and test data for different dataset. The $dev1$ represents the supervised data while $dev2$ comprises the unlabelled-data. }
\vspace{-6mm}
\begin{center}
    \begin{tabular}{ | l | c | c| c|}
\hline
    Data   & LibriSpeech &   TEDLIUM  & Adobe  \\ \hline
    $train$   &  100 hours &  -  &  -  \\
    $dev1$    &    -        & 15 hours & - \\
    $dev2$    &    -        & 50 hours & 20 hours \\
    $dev3$   &   -          & 3 hours  &  - \\
    $test$   &    5 hours    &  2.5 hours & 2.5 hours \\ 
     \hline
    \end{tabular}
\label{table:data_all}
\end{center}
\vspace{-10mm}
\end{table}

\subsection{LibriSpeech}
\label{sec:source_domain} 
\vspace{-2mm}
 We selected 100 hours of LibriSpeech clean portion as the source-domain data, denoted as $train$ in Table~\ref{table:data_all}~\cite{panayotov2015librispeech}. The LibriSpeech test set comprises five hours of clean speech recordings.   
\vspace{-3mm} 

\subsection{TEDLIUM}
\vspace{-2mm}
Experiments are conducted in TEDLIUM speech dataset as the target domain data~\cite{rousseau2012ted}. For our experiments, we used only 15 hours of labelled data ($dev1$ part from Table~\ref{table:data_all}). Furthermore, we use 3 hours and 50 hours of data as the cross validation and unsupervised set ($dev3$ from Table~\ref{table:data_all}) respectively. The test data consists of 2.5 hours. The details of the TEDLIUM corpora can be found in~\cite{rousseau2012ted}.  
\vspace{-2mm}
\subsection{Adobe}
\vspace{-2mm}
The experiments are also conducted on Adobe's internal speech dataset. This corpora contains users uttering a list of commands. An example of a command is , "Move the table". The unsupervised data  contains $\sim$24\,k utterances spoken by 250 speakers with average duration of each utterance being 3\,s ($dev2$). The test data ($test$) consists of 1300 utterances spoken by 50 speakers. The performances of the ASR are reported in terms of word error rate (WER).





\vspace{-2mm}
\subsection{chain model}
\label{sec:chain_model_experiments} 
\vspace{-2mm}
For chain model, 40 dimensional mel frequency cepstral coefficients (MFCC) are extracted from the speech utterance as input features to the neural network~\cite{povey2016purely}. Furthermore, we also use online i-vector features as input. The dimension of the i-vector extractor is fixed to 100.  The DNN uses 7 hidden layers of time delay neural network (TDNN) with 1\,k dimensional units. The DNN is trained to predict senones as the output and LF-MMI is applied as the optimization criteria. The chain model employs a 3-gram LM during decoding phase. The pronunciation dictionary was created on the publicly available CMU-dictionary and include vocabularies from the  training text of LibriSpeech and TEDLIUM datasets.
 \vspace{-2mm}
\subsection{End-to-end ASR}
\vspace{-2mm}
For the end-to-end ASR, 40 dimensional filter-bank energies  are extracted from the utterances to constitute features for the DNN~\cite{watanabe2018espnet, hori2018end}. Delta filter-bank energies and pitch features are appended to the original features to make it 83 dimensional vectors. The end-to-end ASR as described in Section~\ref{sec:end_end} is trained to predict English characters (including semicolon, commas, etc. to make output dimension of 30). The end-to-end ASR uses 3 CNN layers followed by 2 BLSTM layers as the encoder, with the dimension of each layer fixed to 512. The decoder network employs 2 LSTM layers, each with 512 dimensional units. A word based language model is also trained  with a vocabulary size of 50\,k words. The LM uses 2 LSTM layers with dimension of each layers being 1\,k. The training text-data from LibriSpeech and TEDLIUM are used to train the word-based LM.  

\section{Results}
\label{sec:results}
In this section, the results of the end-to-end ASR are presented. The following ASR systems will be analyzed:

\begin{itemize}
    \item $\textbf{LF-MMI}$: This ASR refers to the traditional chain model using LF-MMI optimization criteria. The system is  described in Section~\ref{sec:chain_model_experiments}. This system is trained using the standard kaldi's recipe~\cite{povey2011kaldi}. 
    
    \item $\textbf{End-to-end}$: This refers to end-to-end ASR as presented in Section~\ref{sec:end_end}. The end-to-end ASR is trained to predict characters and referred to as $\textbf{E2E}$. We also trained an end-to-end ASR using a dropout value of 0.2. The dropout is applied to all the layers in encoder and  decoder. The network (with dropout) is trained to predict characters. The end-to-end ASR (source domain data of Section~\ref{sec:source_domain}) with dropout is referred to as $\textbf{E2E}$-$\textbf{drop}$.
    \item $\textbf{Adapted ASR}$: The end-to-end and $\textbf{LF-MMI}$ based ASR are adapted to TEDLIUM labelled data. The end-to-end adapted ASR that is trained in a supervised manner (on $dev1$ data of Table~\ref{table:data_all}) is referred to as $\textbf{E2E}$$_{\textbf{S}}$, while the adapted $\textbf{LF-MMI}$ is referred to as $\textbf{LF-MMI}_{\textbf{S}}$. The end-to-end ASR that exploits the unsupervised data from TEDLIUM is referred to as   $\textbf{E2E}^\textbf{TED}_{\textbf{S+U}}$ and the ASR that uses unlabelled Adobe's internal data is referred to as  $\textbf{E2E}^\textbf{Ab}_{\textbf{U}}$.

    
    
\end{itemize}

\vspace{-4mm}
\subsection{Baseline}

For training the models on source domain, subset of LibriSpeech data is used as described in the Section~\ref{sec:source_domain}. The results of experiments on LibriSpeech clean test set (column 2) are tabulated in Table~\ref{table:results_source_domain}. We observe that $\textbf{LF-MMI}$ outperforms the end-to-end ASR. Furthermore, we also observe that the $\textbf{E2E}$-$\textbf{drop}$ performs worse than $\textbf{E2E}$ by 0.8\% absolute WER. The poor performance of the end-to-end ASR could be due to the limited amount of training data. 

\begin {table}[t!]
\caption{Performance of the various baseline ASR systems in terms of WER (\%) on LibriSpeech clean, TEDLIUM and Adobe test set with the source domain model. The chain model, $\textbf{LF-MMI}$, performs better than the end-to-end ASR systems.}
\vspace{-7mm}
\begin{center}  
    \begin{tabular}{ | l | c| c| c|}
\hline
    Systems   &   LibriSpeech  &  TEDLIUM  &  Adobe  \\ \hline
  $\textbf{LF-MMI}$  &  $\mathbf{7.8}$ & $\mathbf{19.5}$ & $\textbf{29.7}$ \\
  $\textbf{E2E}$ &  11.2 & 38.5 & 40.5\\
 $\textbf{E2E}$-$\textbf{drop}$  & 12.0 & 39.1 & 40.7 \\
     \hline
    \end{tabular} 
\label{table:results_source_domain}

\end{center}
\vspace{-10mm} 
\end{table}

\vspace{-2mm} 
\subsection{Experiments on the TEDLIUM data}
 
The performances of the various source domain ASRs on the TEDLIUM test portion are shown in Table~\ref{table:results_source_domain} (Column 3). It can be observed that the $\textbf{LF-MMI}$ performs the best on the TEDLIUM test set as well. These ASR systems are then adapted to labelled data, $dev1$ (Table~\ref{table:data_all}). For the $\textbf{LF-MMI}_{\textbf{S}}$, retraining all the parameters of the model provides good performance while for end-to-end ASR, retraining the encoder performs the best. From Table~\ref{table:adapt_system_threshold}, it can be observed that $\textbf{LF-MMI}_{\textbf{S}}$ outperforms the  $\textbf{E2E}_{\textbf{S}}$. The end-to-end models are used as seed-model, for exploiting the unsupervised data. We first present results of an experiment employing decoding-score for end-to-end ASR.


\subsubsection{Decoding-score} 
\label{sec:exp_decoding_scores} 
\begin{figure}[t!]
\vspace{-4mm}
    \centering
    \includegraphics[width=\linewidth]{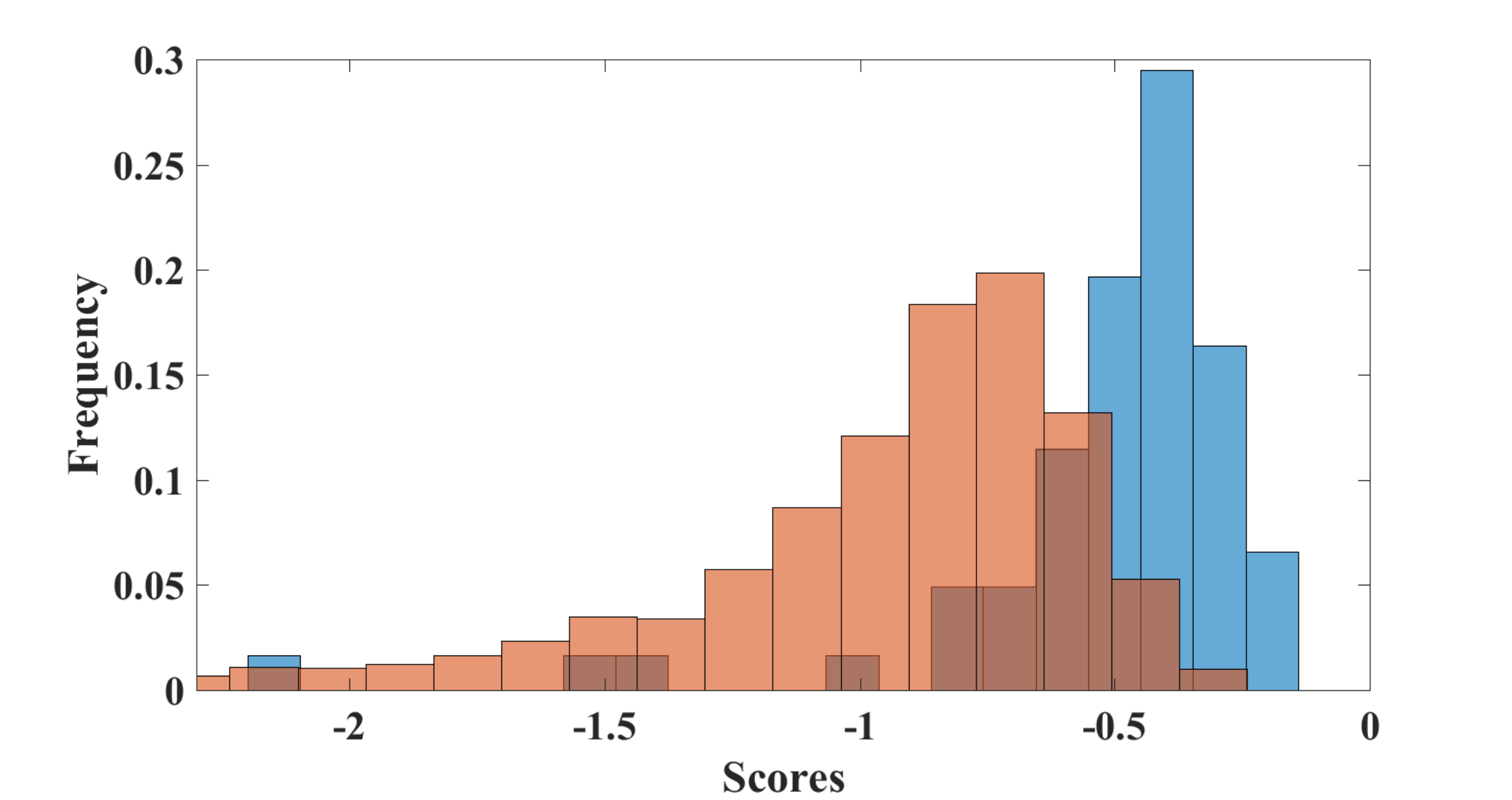}

    \caption{Histogram plot of decoding-scores for set of utterances, with WER $\leq$ 10\% (blue) and  decoding-scores for utterances with WER $>$ 10\% (orange).}
    \label{fig:wer_plot}
     \vspace{-4mm} 
\end{figure}

\vspace{-2mm}
The first step in data-selection process is to fix a threshold on decoding-score generated by the seed-model on unlabelled data. 
The threshold is obtained by minimizing false alarm and miss detection rate as follows. The cross validation (CV) part of TEDLIUM data ($dev3$  data as described in Table~\ref{table:data_all}) is first decoded using the $\textbf{E2E}_\textbf{S}$. For each utterance, the WER is computed. Thus, WER and decoding-score are associated to each utterance. We divide the CV set into two parts, (i) Set1: utterances for which WER $\leq$ $10$\%, and (ii) Set2: WER for these utterances $>$ $10$\%. The histogram plot of the decoding-scores of Set1 and Set2 is illustrated in Figure~\ref{fig:wer_plot}. To  minimize false positive and miss detection (from Figure~\ref{fig:wer_plot}), the threshold should be fixed between -0.3 to -0.6 (Refer to Figure~\ref{fig:wer_plot}). This threshold is applied on the unsupervised data ($dev2$ of Table~\ref{table:data_all}) for selecting speech utterances. 

The end-to-end ASR (seed-model) is applied to decode the $dev2$ data to obtain text-transcripts (10-best) and decoding-scores for an utterance. Data-selection is applied on these decoding-scores for filtering the highly confident outputs (for further retraining the system). The results of data-selection process using two threshold values are illustrated in Table~\ref{table:adapt_system_threshold}. We observed that  10-best hypothesis is beneficial for $\textbf{E2E}^\textbf{TED}_{\textbf{S+U}}$  than 1-best hypothesis. Data-selection based on 1-best hypothesis provides WER of 29.5\% while data-selection using 10-best decoded-outputs provides WER of 28.9\%. Furthermore, the performance of $\textbf{E2E}^\textbf{TED}_{\textbf{S+U}}$ does not improve on applying more than 10-best hypothesis. From Table~\ref{table:adapt_system_threshold}, it can be observed that  $\textbf{E2E}^\textbf{TED}_{\textbf{S+U}}$ performs better with threshold of -0.5. We observed that thresholds less than -0.3 lead to the selection of shorter duration utterances. The $\textbf{E2E}^\textbf{TED}_{\textbf{S+U}}$ outperforms  $\textbf{E2E}_{\textbf{S}}$ by 0.3\% absolute WER. In rest of the experimental section, threshold of -0.5 is applied on decoding-score for data-selection.

We augment the $N$-best hypothesized transcripts (generated by $\textbf{E2E}$$_{\textbf{S}}$) by using dropout mechanism from the  $\textbf{E2E}$-$\textbf{drop}$. The process for data-selection is described in Section~\ref{sec:semi_supervised}. From Table~\ref{table:decoding_score_selection}, it can be observed that significant gain in performance is obtained by  this approach ($\textbf{E2E}^\textbf{TED}_{\textbf{S+U}}$ + $\textbf{E2E}$-$\textbf{drop}$). Furthermore, this technique outperforms  the $\textbf{E2E}$$_{\textbf{S}}$ by 2\% absolute WER (29.2\% to 27.2\%) showing the importance of localizing the uncertainties in the model. 
\vspace{-3mm}
\subsubsection{Entropy} 
\vspace{-2mm}
We also explore an approach of using entropy as data-selection criteria as described in Section~\ref{sec:semi_supervised}. From Table~\ref{table:decoding_score_selection}, it can be observed that the performance of $\textbf{E2E}^\textbf{TED}_{\textbf{S+U}}$  does not improve upon $\textbf{E2E}_{\textbf{S}}$ on using additional unsupervised data. Furthermore, we apply dropout mechanism as described in Section~\ref{sec:semi_supervised} for augmenting the data (as done in Section~\ref{sec:exp_decoding_scores} on decoding-score). It can be observed that this approach outperforms $\textbf{E2E}_{\textbf{S}}$ by 0.4\% absolute WER.

\vspace{-3mm}
\subsection{Experiments on Adobe's internal dataset} 
\vspace{-2mm}
We performed the data-selection process using decoding-score and entropy based confidence scores on Adobe's internal data as well. Due to the lack of development data from Adobe, the thresholds from TEDLIUM are used for data-selection process. For example, threshold of -0.5 is applied on decoding-score for utterance-selection. The result of data-selection process using decoding-score is shown in Table~\ref{table:decoding_score_selection}. From the table, it can be observed that $\textbf{E2E}^\textbf{Ab}_{\textbf{U}}$ performs best (retrained with 10-best hypothesized transcripts) with a WER of 32.2\%. Furthermore,  additional transcripts generated by the dropout model are used for augmenting the 10-best hypothesized text. The performance of this approach ($\textbf{E2E}^\textbf{Ab}_{\textbf{U}}$ + $\textbf{E2E}$-$\textbf{drop}$) does not improve upon the result of 32.2\% WER. The poor performance could be due to choice of non optimal threshold on decoding-score for data-selection process. It can be observed that  dropout mechanism benefits performance of the $\textbf{E2E}^\textbf{Ab}_{\textbf{U}}$ + $\textbf{E2E}$-$\textbf{drop}$ over the non-adapted $\textbf{E2E}$ using entropy based data-selection criteria.



\begin {table}[t!]
\caption{Performance of various ASR systems in terms of WER (\%) on TEDLIUM test set. The details of supervised and unsupervised data (adapt-data) have been tabulated in Table~\ref{table:data_all}.}
\vspace{-4mm} 
\begin{center}  
    \begin{tabular}{ | l | c | c| c|}
\hline
    Systems   &  adapt-data    & Threshold & TEDLIUM \\ \hline
   $\textbf{LF-MMI}_{\textbf{S}}$ & $dev1$  & - &     $\mathbf{18.5}$ \\ 
   $\textbf{E2E}$$_{\textbf{S}}$ & $dev1$  & - &     29.2 \\ 
    $\textbf{E2E}^\textbf{TED}_{\textbf{S+U}}$ & $dev1$+$dev2$  &  -0.3 &   29.7 \\
    $\textbf{E2E}^\textbf{TED}_{\textbf{S+U}}$  & $dev1$+$dev2$   & -0.5 &    28.9 \\
     \hline
    \end{tabular} 
\label{table:adapt_system_threshold}
\end{center}
\vspace{-3mm} 
\end{table}

\begin {table}[t!]
\caption{Performance of various ASR systems on TEDLIUM (TED) and Adobe (Adb) test set  using decoding-score (dec-score)  and entropy based data-selection (data-sel) criteria. The $\textbf{E2E}^\textbf{TED}_{\textbf{S+U}}$ + $\textbf{E2E}$-$\textbf{drop}$  performs the best on TEDLIUM test set.}
\vspace{-4mm}
\begin{center}  
    \begin{tabular}{ | l | c |c| c|}
\hline
    Systems   &   $data$-$sel$ & TED  & Adb  \\ \hline
   $\textbf{E2E}$   &  - &  38.5 & 40.5 \\
   $\textbf{E2E}$$_{\textbf{S}}$ & - &  29.2  &  -  \\
   $\textbf{E2E}^\textbf{TED}_{\textbf{S+U}}$   &  $dec$-$score$ &  28.9  & -  \\ 
      $\textbf{E2E}^\textbf{Ab}_{\textbf{U}}$   & $dec$-$score$ &  - &  $\mathbf{32.2}$ \\   
    $\textbf{E2E}^\textbf{TED}_{\textbf{S+U}}$ + $\textbf{E2E}$-$\textbf{drop}$ & $dec$-$score$  & $\mathbf{27.2}$   &  - \\
    $\textbf{E2E}^\textbf{Ab}_{\textbf{U}}$ + $\textbf{E2E}$-$\textbf{drop}$ & $dec$-$score$  &  -  & 34.3 \\ 
   $\textbf{E2E}^\textbf{TED}_{\textbf{S+U}}$   & $entropy$  &   29.2  &  - \\
   $\textbf{E2E}^\textbf{Ab}_{\textbf{U}}$    & $entropy$  &   -  &  38.1 \\
  $\textbf{E2E}^\textbf{TED}_{\textbf{S+U}}$ + $\textbf{E2E}$-$\textbf{drop}$   & $entropy$  &   28.8  &  - \\  
 $\textbf{E2E}^\textbf{Ab}_{\textbf{U}}$ + $\textbf{E2E}$-$\textbf{drop}$ &  $entropy$  &   -  &  37.7 \\ 
  
    
     \hline
    \end{tabular} 
\label{table:decoding_score_selection}
\end{center}
\vspace{-3mm} 

\end{table}

\normalsize 



\section{Conclusions}
\label{sec:conclusions} 

Techniques for semi-supervised learning for ASR were investigated in this paper. For exploiting unlabelled data, the baseline system employs a single best hypothesized  text-transcript. As opposed to this approach, we proposed to capture the uncertainties by applying a dropout mechanism to generate multiple hypothesized transcripts. Furthermore, we also used data-selection mechanism to filter the highly confident hypotheses. The techniques were evaluated in publicly available TEDLIUM and Adobe's internal dataset. Experiments show that the proposed approach ($\textbf{E2E}^\textbf{TED}_{\textbf{S+U}}$ + $\textbf{E2E}$-$\textbf{drop}$) outperforms the baseline method by 2\% absolute reduction in WER on TEDLIUM test set. 


\section{Acknowledgements}

This work was done under the ``SM2 - Extracting semantic meaning from spoken material" project, partially supported by the Swiss Innovation Agency (InnoSuisse) as well as by a research grant from Adobe Research, USA.

\bibliographystyle{IEEEtran}

\bibliography{mybib}


\end{document}